\documentclass[useAMS,usenatbib]{mn2e}

\newcommand{\ltsima} {$\; \buildrel < \over \sim \;$} 
\newcommand{\simlt}  {\lower.5ex\hbox{\ltsima}}            
\newcommand{\gtsima} {$\; \buildrel > \over \sim \;$} 
\newcommand{\simgt}  {\lower.5ex\hbox{\gtsima}}            

\title[RS~Oph: amplitude of flickering]{Optical flickering  of  the recurrent nova RS~Ophiuchi: 
amplitude$-$flux relation\thanks{Based on observations obtained in National Astronomical Observatory Rozhen and Belogradchik Observatory, Bulgaria.}}
\author[Zamanov  et al.]{R. Zamanov,$^{1}$\thanks{E-mail: rkz@astro.bas.bg (RZ); glatev@astro.bas.bg (GL); kstoyanov@astro.bas.bg (KS)}
G. Latev,$^{1}$
S. Boeva,$^{1}$
J. L. Sokoloski,$^{2}$
K. Stoyanov,$^{1}$
R. Bachev,$^{1}$ 
\newauthor B. Spassov,$^{1}$  
G. Nikolov,$^{1}$
V. Golev$^{3}$
and  S. Ibryamov$^{1}$
\vspace{0.3cm}
\\
$^{1}$Institute of Astronomy and National Astronomical Observatory, Bulgarian Academy of Sciences, Tsarigradsko Shose No. 72, \\
1784 Sofia, Bulgaria
\vspace{0.1cm} \\
$^{2}$ Columbia Astrophysics Laboratory, Columbia University, 550 West 120th Street, New York, NY 10027, USA 
\vspace{0.1cm} \\
$^{3}$ Department of Astronomy, Faculty of Physics, St Kliment Ohridski University of Sofia, 5 James Bourchier Blvd., 1164, Sofia, Bulgaria
}
\begin{document}

\date{Accepted, 2015 April 17. Received 2015 April 17; in original form 2015 January 7}

\pagerange{\pageref{firstpage}--\pageref{lastpage}} \pubyear{2014}

\maketitle

\label{firstpage}

\begin{abstract}

We report observations of the flickering variability of the symbiotic
recurrent nova RS~Oph at quiescence in five bands ($UBVRI$).  We find
evidence of a correlation between the peak-to-peak flickering amplitude
($\Delta F$) and the average flux of the hot component ($F_{\rm av}$). The
correlation is highly significant, with a correlation coefficient of 0.85
and a $p$-value of~$\sim 10^{-20}$. Combining the data from all wavebands,
we find a dependence of the type $\Delta F \propto F^k_{\rm av}$, with
power-law index $k = 1.02 \pm 0.04$ for the $UBVRI$ flickering of RS~Oph.
Thus, the relationship between the 
amplitude of variability  and the 
average flux of the hot component is consistent with linearity.
The rms amplitude of flickering is on average 8 per cent  ($\pm2$ per cent) of $F_{\rm av}$.
The detected correlation is similar to that found in accreting black holes/neutron stars 
and cataclysmic variables. 
The possible reasons are briefly discussed. The data are
available upon request from the authors. 
\end{abstract}


\section{Introduction}

In the symbiotic recurrent nova RS~Ophiuchi (HD 162214),
a near-Chandrasekhar-mass white dwarf (WD) accretes material from
a red giant companion (e.g., Hachisu \& Kato 2001; Sokoloski et
al. 2006; Bode 2010 and references therein).  
It experiences nova eruptions approximately every 20 yr. RS~Oph 
has undergone recorded outbursts in  
1898, 1933, 1958, 1967, and 1985 (Rosino 1987), 
with a possible additional outburst in 1907 (Schaefer 2010).
The most recent eruption occurred on 2006 February 12 (Narumi et al. 2006). 

Using infrared radial velocity measurements, Fekel et al. (2000) found
that RS~Oph has an eccentricity $e \approx 0$ and that the red giant and
WD have masses of 2.3 M$_{\odot}$ and close to 1.4
M$_{\odot}$, respectively, with a separation between the components of $a
= 2.68 \times 10^{13}$ cm.  Brandi et al. (2009), on the basis of optical
and infrared spectra, derived a mass ratio $ q = M_g / M_h = 0.59 \pm
0.05 $ and fine-tuned the orbital period to $453.6 \pm 0.4$ d.

Worters et al. (2007) 
proposed that for the range of spectral types
suggested for the red giant  
in the RS~Oph system, its radius is smaller than its Roche lobe, 
and accretion on to the WD may occur only from the red giant wind.
Wynn (2008) considered that both Roche lobe overflow and stellar wind
capture are plausible methods for the accretion process  
in RS~Oph. He also proposed that the accretion disc is probably a hybrid
type with stable, cold outer regions and stable, hot inner regions
(Wynn 2008). 
$Chandra$ and \textit{XMM}$-$\textit{Newton} observations obtained after
the 2006 outburst suggest that the mass accretion rate at that time
was about $2 \times 10^{-8}$ M$_\odot$~yr$^{-1}$ (Nelson et al.~2011).

Flickering (aperiodic broad-band variability)
is a type of variability observed in the three main 
classes of binaries
that contain WDs  
accreting  material from a companion mass-donor star:  
cataclysmic variables (CVs), supersoft X-ray binaries, 
and symbiotic stars (Sokoloski 2003). 
The flickering is not only observed in accreting WD, but also in 
accreting black holes and neutron stars (e.g. Belloni et al.(2002) and
references therein).  
The flickering of RS~Oph has been detected by Walker (1977), among others. 
Systematic searches  for  flickering variability 
in  symbiotic stars and related objects (Dobrzycka et al. 1996; 
Sokoloski, Bildsten \& Ho \ 2001;  Gromadzki et al. 2006) have shown
that among $\sim200$ known symbiotic stars, only
10 present flickering -- RS~Oph, T~CrB, MWC~560, Z~And, V2116~Oph, CH~Cyg, RT~Cru, {\it o}~Cet, and more recently 
V407~Cyg (Kolotilov et al. 2003) and  V648 Car (Angeloni et al. 2012).  

Here, we present new observations of the flickering variability of
RS~Oph in the $UBVRI$ bands,  
and investigate the behaviour of the  flickering amplitude.

{\footnotesize
\begin{table*} 
\caption{CCD observations of RS~Oph. 
}
\begin{center}
\begin{tabular}{lllrrrrcrrccccrlcr}
\hline
Date         & Telescope     & Band       & \textsc{ut}  & Exp-time  &  N$_{pts}$  &  Average & min--max     & stdev  & err   \\
             &               &            & start--end   & (s)     &             & (mag)    & (mag)-(mag) & (mag)  & (mag) \\
\hline
1997 09 01   & 1 m Lick  & $B$ & 03:41--06:53 & 22 &  233 & 12.297 & 12.078--12.444 &  0.068 &  0.005 & \\
1997 09 02   & 1 m Lick  & $B$ & 03:29--06:49 & 22 &  238 & 12.484 & 12.305--12.653 &  0.086 &  0.005 & \\
1998 07 19   & 1 m Lick  & $B$ & 05:34--08:43 & 60 &  128 & 12.392 & 12.182--12.617 &  0.108 &  0.004 & \\ 
1998 07 20   & 1 m Lick  & $B$ & 05:02--08:38 & 60 &  149 & 12.235 & 12.089--12.389 &  0.071 &  0.003 & \\
1998 07 22   & 1 m Lick  & $B$ & 07:01--08:59 & 60 &   82 & 12.109 & 11.978--12.404 &  0.076 &  0.003 & \\
2008 07 09   &	60 cm Roz	        & $V$ &19:44--21:08&   60     &  50  &  11.218  & 11.313--11.110 & 0.052 &	0.020   &  \\  
2009 06 14   &	60 cm Roz	        & $B$ &23:35--00:32&	30     &  70  &  11.722  & 11.605--11.855 & 0.060 &	0.020   &  \\  
2009 06 14   &	60 cm Roz	        & $I$ &23:36--00:32&	 5     &  70  &   8.852	 &  8.773--8.951 & 0.032 &   0.010   &  \\  
2009 06 14   &	60 cm Bel                & $V$ &23:37--00:43&	20     &  95  &  10.703  & 10.557--10.817 & 0.049 & 	0.005   &  \\  
2009 06 14   &	60 cm Bel                & $R$ &23:37--00:43&	10     &  95  &   9.839  &  9.701--9.915 & 0.040 &   0.005   &  \\  
2009 06 15   &	60 cm Roz	        & $B$ &22:24--00:03& 	40     &  81  &  11.858  & 11.714--12.045 & 0.062 &   0.010   &  \\  
2009 06 15   &	60 cm Roz	        & $I$ &22:25--00:04& 	10     &  81  &   8.955	 &  8.848--9.037 & 0.035 &   0.005   &  \\  
2009 06 15   &	60 cm Bel                & $V$ &23:31--01:02&	20     & 130  &  10.795  & 10.693--10.934 & 0.046 &	0.005   &  \\  
2009 06 15   &	60 cm Bel                & $R$ &23:31--01:02&	10     & 130  &   9.926  &  9.834--10.038 & 0.038 &   0.005   &  \\  
2010 04 30   &  60 cm Roz                & $B$ &22:47--00:19&  30,60   &  65  & 11.737  & 11.603--11.848 & 0.059 &  0.006 &  \\   
2010 04 30   &  60 cm Roz	        & $V$ &22:48--00:21&  10      &  65  & 10.621  & 10.521--10.718 & 0.048 &  0.006 &  \\   
2010 05 01   &  60 cm Roz                & $B$ &22:19--00:19&  60      &  68  & 11.431  & 11.121--11.612 & 0.128 &  0.004 &  \\   
2010 05 01   &  60 cm Roz	        & $V$ &22:19--00:20&  20      &  69  & 10.375  & 10.093--10.541 & 0.113 &  0.003 &  \\   
2010 05 02   &  60 cm Roz	        & $U$ &22:42--00:25&  120     &  30  & 11.576  & 11.354--11.739 & 0.118 &  0.013 &  \\   
2010 05 02   &  60 cm Roz                & $B$ &22:56--00:27&  60      &  29  & 11.643  & 11.457--11.777 & 0.097 &  0.006 &  \\   
2012 04 27   &  70 cm Sch                & $U$ &00:07--01:34& 60,120   &  54  &  12.037 & 11.850--12.244 & 0.086 &  0.011 &  \\   
2012 04 27   &  60 cm Roz                & $B$ &00:28--01:43&  60      &  61  &  12.112 & 11.934--12.271 & 0.070 &  0.006 &  \\   
2012 04 27   &  60 cm Bel                & $V$ &00:03--01:39&  20      & 141  &  10.982 & 10.876--11.120 & 0.052 &  0.006 &  \\   
2012 04 27   &  60 cm Roz                & $I$ &00:29--01:44&   3      &  61  &   9.044 &  8.986--9.110 & 0.029 &  0.006 &  \\   
2012 06 13   &  60 cm Roz                & $U$ &21:45--23:36& 120      &  33  &  12.599 & 12.407--12.762 & 0.089 &  0.014 &  \\   
2012 06 13   &  60 cm Roz                & $B$ &21:47--23:37&  20      &  34  &  12.531 & 12.380--12.671 & 0.086 &  0.012 &  \\   
2012 06 13   &  60 cm Roz                & $V$ &21:43--23:37&  10      &  34  &  11.376 & 11.274--11.492 & 0.063 &  0.007 &  \\   
2012 06 13   &  60 cm Roz                & $R$ &21:43--23:38&   5      &  34  &  10.363 & 10.282--10.451 & 0.050 &  0.002 &  \\   
2012 06 13   &  60 cm Roz	        & $I$ &21:44--23:38&   3      &  33  &   9.266 &  9.172--9.342 & 0.037 &  0.005 &  \\   
2012 07 18   &  2.0 m Roz  	        & $U$ &21:02--22:54& 300      &  21  &  12.673 & 12.567--12.814 & 0.065 &  0.008 &  \\   
2012 07 18   &  70 cm Sch  	        & $B$ &20:45--23:00&  20      & 317  &  12.722 & 12.614--12.851 & 0.046 &  0.010 &  \\   
2012 07 18   &  2 m Roz  	        & $V$ &21:02--22:58&  15      & 200  &  11.540 & 11.462--11.617 & 0.030 &  0.006 &  \\   
2012 07 21   &  60 cm Bel                & $B$ &18:59--22:42&  30      & 219  &  12.593 & 12.377--12.794 & 0.073 &  0.014 &  \\   
2012 07 21   &  60 cm Bel                & $V$ &18:59--22:43&  10      & 222  &  11.432 & 11.288--11.628 & 0.050 &  0.009 &  \\   
2012 07 21   &  60 cm Bel                & $R$ &19:00--22:43&   5      & 222  &  10.422 & 10.257--10.524 & 0.038 &  0.006 &  \\   
2012 07 21   &  60 cm Bel                & $I$ &19:00--22:43&   3      & 222  &   9.326 &  9.205--9.439 & 0.054 &  0.005 &  \\   
2012 08 15   &  60 cm Roz                & $B$ &18:41--21:10&  60      &  83  &  12.904 & 12.618--13.208 & 0.216 &  0.009 &  \\   
2012 08 15   &  60 cm Roz                & $R$ &18:42--21:10&   5      &  83  &  10.579 & 10.399--10.730 & 0.116 &  0.008 &  \\   
2012 08 15   &	60 cm Roz                & $I$ &18:42--21:10&   5      &  83  &   9.404 &  9.301--9.501 & 0.065 &  0.006 &  \\   
2012 08 16   &	2.0 m Roz                & $U$ &18:48--20:51&  180,240 &  28  &  12.782 & 12.659--12.904 & 0.070 &  0.022 &  \\   
2012 08 16   &	2.0 m Roz                & $V$ &18:47--20:51&  10      & 224  &  11.563 & 11.472--11.633 & 0.037 &  0.004 &  \\   
2012 08 16   &  70 cm Sch                & $B$ &18:35--20:44&  30      & 224  &  12.769 & 12.561--12.897 & 0.071 &  0.007 &  \\   
2013 07 02   &  60 cm Roz                & $B$ &20:41--22:43&   60     &  46  & 12.196  & 11.982--12.368 & 0.074 & 0.008  &  \\   
2013 07 02   &  60 cm Roz                & $V$ &20:42--22:44&   60     &  49  & 11.115  & 10.988--11.237 & 0.060 & 0.005  &  \\   
2013 07 10   &  70 cm Sch                & $B$ &21:23--23:42&   30     & 126  &  12.033 & 11.887--12.207 & 0.069 & 0.008  &  \\   
2013 07 10   &  70 cm Sch               & $V$ &21:23--23:43&   15     & 128  &  11.041 & 10.896--11.196 & 0.065 & 0.006  &  \\   
2013 08 12   &  2.0 m  Roz               & $U$ &19:09--21:27&  180     &  34  &  11.996 & 11.766--12.171 & 0.121 & 0.009  &  \\   
2013 08 12   &  70 cm Sch                & $B$ &18:50--21:30& 15,20    & 470  &  12.129 & 11.916--12.270 & 0.083 & 0.012  &  \\   
2013 08 12   &  2.0m Roz                & $V$ &19:12--21:30&   3, 5   & 335  &  11.060 & 10.897--11.203 & 0.069 & 0.005  &  \\   
2013 08 12   &  60 cm Roz                & $R$ &19:17--21:29&   3      & 285  &  10.129 &  9.985--10.240 & 0.057 & 0.007  &  \\   
2013 08 12   &  60 cm Roz                & $I$ &19:17--21:28&   3      & 285  &   9.069 &  8.954--9.185  & 0.046 & 0.004  &  \\   
2013 08 13   &  2.0 m Roz                & $U$ &18:58--21:21&   180    &  37  &  12.487 & 12.296--12.658 & 0.101 & 0.013  &  \\   
2013 08 13   &  70Sch+60Roz             & $B$ &20:40--21:22&    20    &  92  &  12.579 & 12.476--12.659 & 0.042 & 0.010  &  \\   
2013 08 13   &  2.0 m Roz                & $V$ &18:58--21:22&     3    & 362  &  11.391 & 11.274--11.503 & 0.060 & 0.004  &  \\   
2013 08 13   &  60 cm Roz                & $R$ &18:48--21:21&     3    & 320  &  10.365 & 10.265--10.467 & 0.046 & 0.008  &  \\   
2013 08 13   &  60 cm Roz                & $I$ &18:48--21:21&     3    & 320  &   9.197 &  9.107--9.288  & 0.038 & 0.005  &  \\   
2013 09 05   &  60 cm Bel                & $B$ &18:45--19:47&    30    &  52  &  12.033 & 11.864--12.226 & 0.086 & 0.015  &  \\ 
2013 09 05   &  60 cm Bel                & $V$ &18:46--19:47&    10    &  51  &  10.866 & 10.730--11.028 & 0.074 & 0.010  &  \\ 
2013 09 05   &  60 cm Bel                & $R$ &18:46--19:47&     3    &  51  &   9.933	&  9.819--10.082 & 0.067 & 0.015  &  \\ 
2013 09 05   &  60 cm Bel                & $I$ &18:46--19:48&     3    &  51  &   8.983 &  8.875--9.091 & 0.052 & 0.015  &  \\
2013 09 06   & 60 cm Bel 	        & $B$ &18:41--19:25& 30       &  39  &  12.012 & 11.848--12.152 & 0.090 & 0.006  &  \\   
2013 09 06   & 60 cm Bel 	        & $V$ &18:41--19:25& 10       &  39  &  10.865 & 10.723--10.977 & 0.071 & 0.004  &  \\   
2013 09 06   & 60 cm Bel 	        & $R$ &18:42--19:25&  5       &  39  &   9.889 &  9.786--9.960  & 0.047 & 0.003  &  \\   
2013 09 06   & 60 cm Bel 	        & $I$ &18:42--19:25&  5       &  39  &   8.971 &  8.868--9.047  & 0.043 & 0.003  &  \\   
\hline																     
\end{tabular}															     
\end{center}															     
\label{Tab2}															     
\end{table*}		
}													     
																     

 \begin{figure*}    
   \vspace{6.7cm}     
   \includegraphics{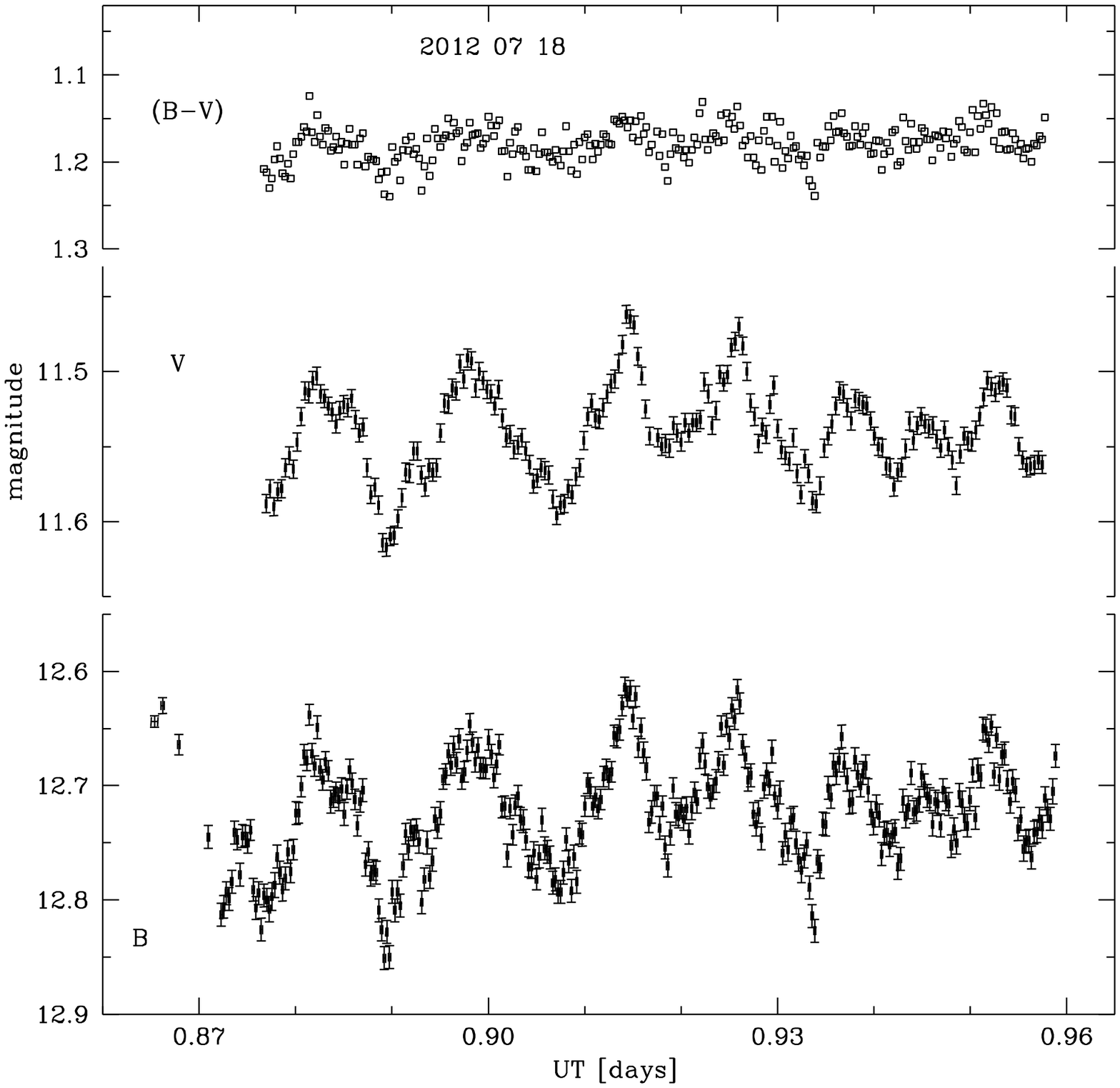}      
   \includegraphics{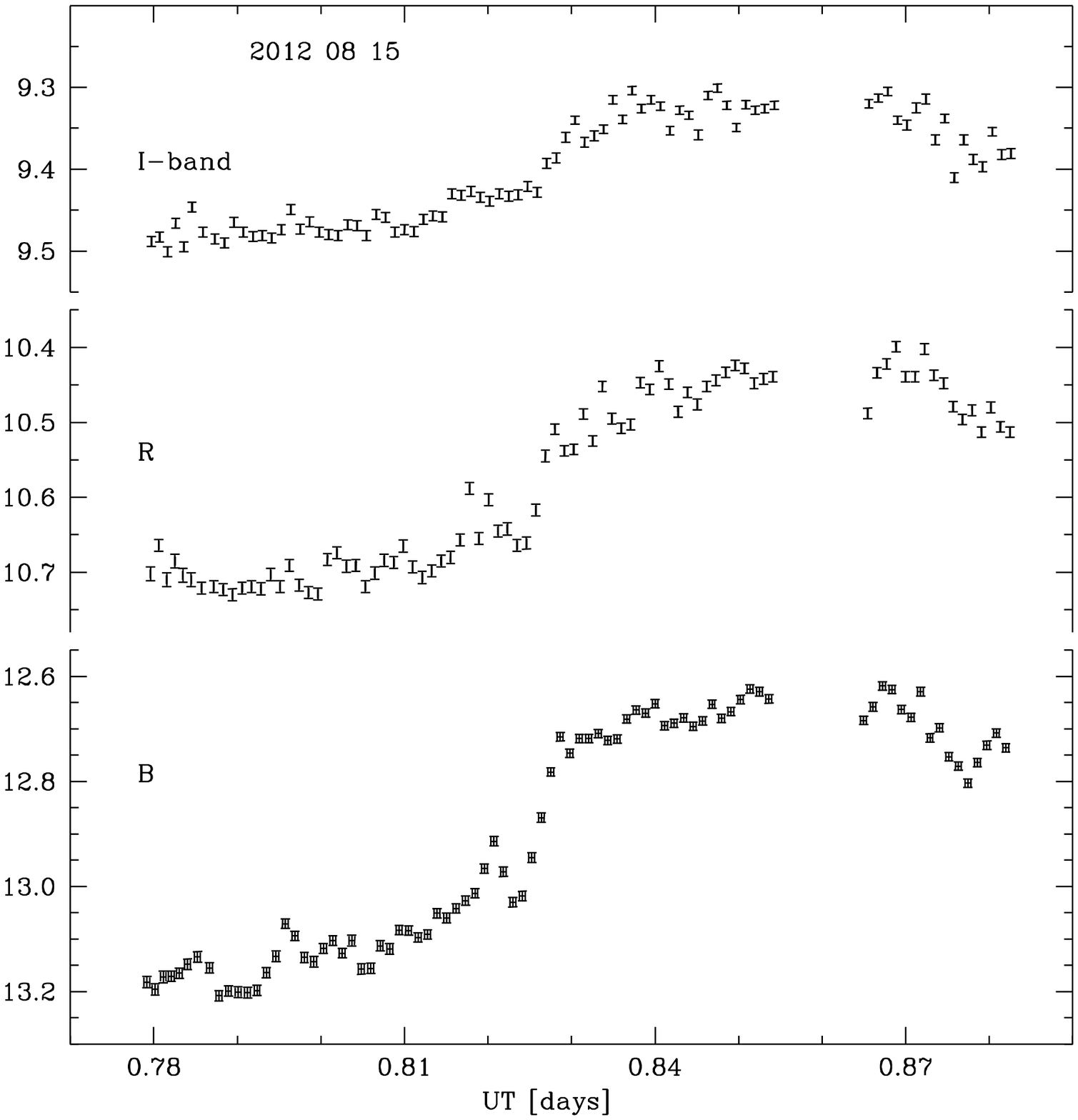}      
   \includegraphics{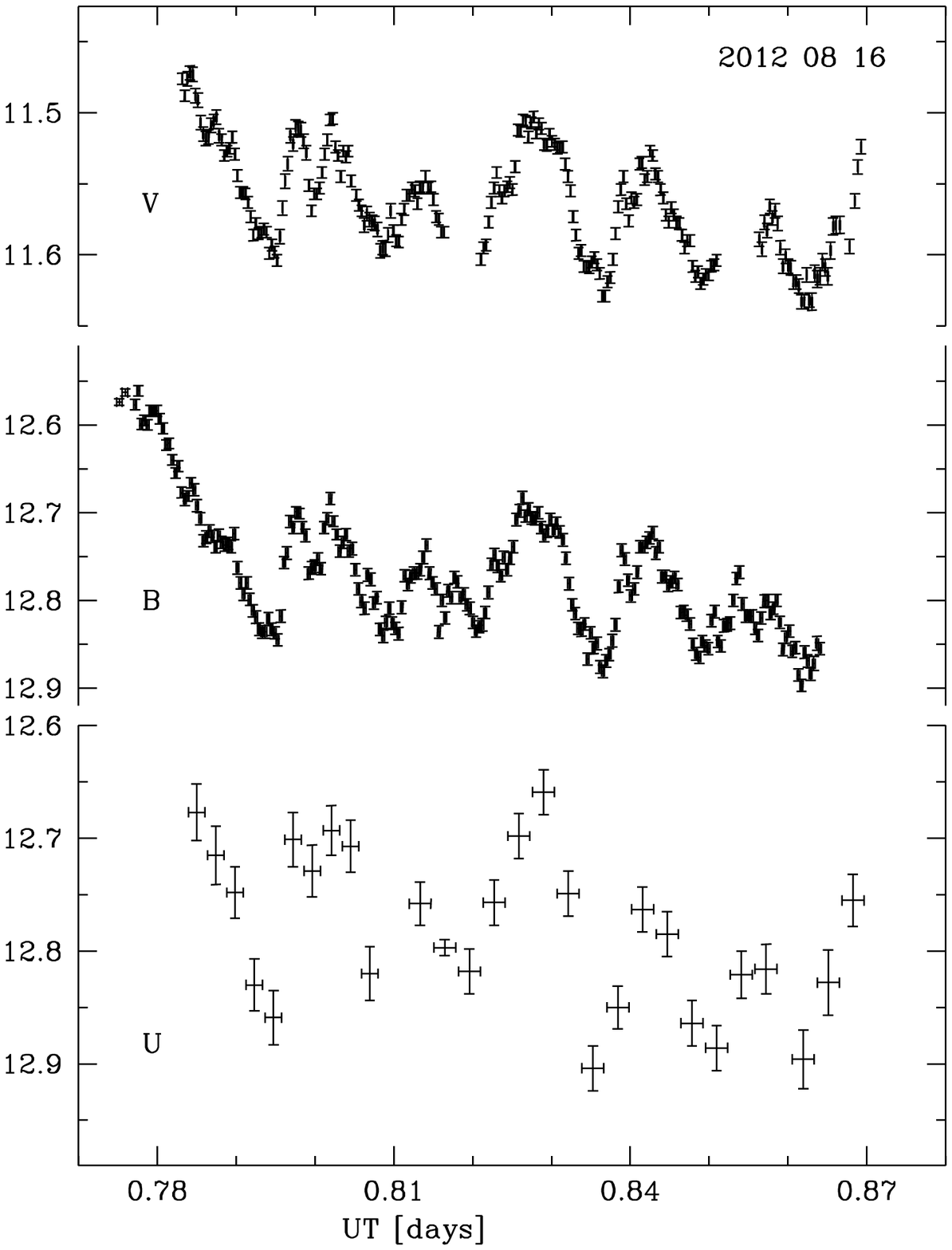}      
   \caption[]{Example light curves of RS~Oph. }
   \label{fig.example}      
 \end{figure*}	     

\section[]{Observations}

Although some of the observations presented here were performed before the 2006
outburst using the 1.0 m Nickel  
telescope at UCO/Lick Observatory on Mt. Hamilton near San Jose, CA (USA), 
the majority were obtained 
between 2008 July and 2013 September with 
the 2 m RCC telescope, the 50/70 cm Schmidt telescope, the 60 cm  telescope
of the Bulgarian National Astronomical Observatory  Rozhen, 
and the 60 cm telescope of the Belogradchick Astronomical
Observatory. All of the 
telescopes are equipped with CCD cameras. The 2 m RCC telescope
is equipped with a dual channel focal reducer (Jockers et al.  2000) and 
can observe simultaneously in two bands -- $U$ (blue
channel) and $V$ (red channel).  
 
All of the CCD images have been bias subtracted and flat fielded,
and standard aperture photometry has been performed. The data from
Lick Observatory were reduced using \textsc{idl}.  For the data from Rozhen and
Belogradchick, 
the reduction and aperture photometry were done with \textsc{iraf} and
checked with alternative software packages. 
Depending upon the field of view, we used between two and six comparison
stars from the list of Henden \& Munari (2006).  
Table~\ref{Tab2} 
lists the date in format YYYY MMM DD, the telescope, band, \textsc{ut}-start,
and \textsc{ut}-end of the run, exposure time, number of  
CCD images obtained, average magnitude in the corresponding band, 
minimum--maximum magnitudes in each band, standard
deviation of the mean, and typical observational error.
The exposure times were from 3 to 240 s, and the read-out-times
were from 3 to 20 s, depending on the brightness of the object
and the observational setup.  
The typical accuracy of the photometry was $\sim 0.01$~mag (see
Table~\ref{Tab2} for the accuracy of the individual light curves). 

In addition to the new observations, we also used published data
(Zamanov et al. 2010;  Zamanov \& Bachev 2007). 
A few examples of our observations are presented in
Fig.~\ref{fig.example}.

\section{Red giant contribution}  

Pavlenko et al. (2008) modelled a 2006 August spectrum of RS~Oph in
the 1.4--2.5 $\mu m$ range and determined the following parameters
for the red giant: $T_{\rm eff} = 4100 \pm 100$~K, $\log g = 0.0 \pm 0.5$,
$[Fe/H] = 0.0 \pm 0.5$, $[C/H] = -0.8 \pm 0.2$, and $[N/H] = +0.6 \pm
0.3$. These abundances may vary considerably, however, if the red
giant is contaminated by the nova ejecta, as has been suggested by
Scott et al. (1994). Irradiation of the red giant by the still-hot WD
may also be a complicating factor in the immediate aftermath of an
eruption.  Using near-infrared spectroscopy in
the 1--5 $\mu m$ range, Rushton et al. (2010) found $T_{\rm eff} = 4200 \pm
200$~K for the red giant.

The grids of colours for cool stars (Houdashelt, Bell, Sweigartet
2000) in Johnson--Cousins system (for $\log g = 0.0$, $[Fe/H] = 0.0$) 
give for $T_{\rm eff} =4000$: $U-V=4.178$, $B-V=1.725$, $V-R=0.822$, and
$V-I=1.520$, and for $T_{\rm eff} = 4250$: $U-V=3.731$, $B-V=1.575$,
$V-R=0.723$, and
$V-I=1.325$.  For the red giant of RS~Oph, we 
assume  $(U-V)_0=3.95$,
$(B-V)_0=1.65$, $(V-R)_0$=0.77, and $(V-I)_0=1.42$.

Skopal (2015) modelled the spectral energy distribution of RS~Oph
adopting $V \sim 12.0$ for the red giant.  Using decomposition
of the spectrum of RS~Oph in quiescence, Kelly et al. (2014) estimated
that the giant star should be 0.65 mag fainter in $V$ than the total
magnitude of the binary system, which they suppose to be $\sim 11.5$.
This gives for the giant $V \approx 12.15$.

During the time of our observation, the $V$ brightness of RS~Oph varied
between 10.093 and 11.633  mag. 
However, after the 2006 outburst the brightness of RS~Oph achieved a
minimum value $V \sim 12.25$ (AAVSO data;  
Henden, 2013), which we 
take to be 99\% due to the red giant (because
some contribution  
from the WD and nebula should exist). 
The red giant is `perhaps slightly variable', as noted by Rosino,
Bianchini\& Rafanelli (1982) 
and Rushton et al. (2010). Scott et al. (1994) discussed a mechanism
whereby the outbursts contaminate    
the red giant with excess carbon, which is subsequently
convected away. This would have a negligible influence on the broad-band
optical magnitudes of the red giant, and we 
assume that the red giant is non-variable in $UBVRI$. 
%
%
We 
adopt interstellar reddening towards RS~Oph of $E(B-V) = 0.73$
(Snijders 1987). 


 \begin{figure}    
   \vspace{9.0cm}   
   \includegraphics{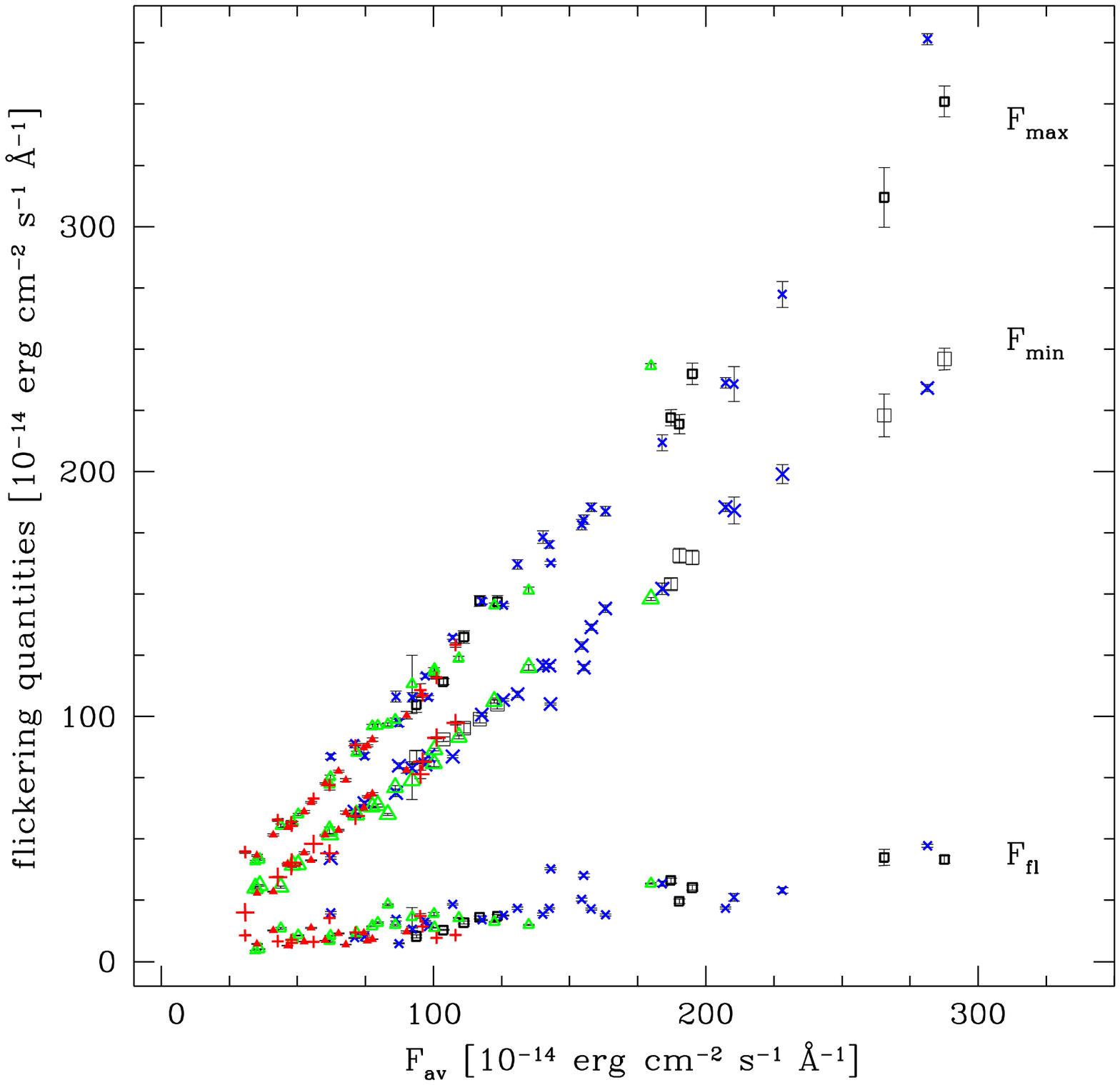}      
   \caption[]{Measured flickering quantities $F_{\rm max}$, $F_{\rm min}$ and
     $F_{\rm fl}$ versus the average flux, F$_{\rm av}$,    
   of the hot component. The symbols are  U -- black squares, B --
   blue crosses, V -- green open triangles, R -- red plusses, I -- red
   filled triangles. 
   The individual error bars are plotted in black. }
   \label{fig.quant}      
 \end{figure}	     

\section{Flickering quantities}  

We converted the magnitudes into fluxes, adopting   
fluxes for a zero-magnitude star of  
F$_{0}(U)= 4.167\times 10^{-9}$,
F$_{0}(B)= 6.601\times 10^{-9}$, 
F$_{0}(V)= 3.610\times 10^{-9}$, 
F$_{0}(R)= 2.256\times 10^{-9}$, and
F$_{0}(I)= 1.226\times 10^{-9}$  $erg \; cm^{-2} s^{-1} \AA^{-1}$
(Bessel 1979).  The observed flux during a given night was corrected   
for the contribution of the red giant and interstellar extinction. 
For each run, we calculate the  following dereddened quantities: \\
$F_{\rm max}$ -- the maximum flux of the hot component; \\
$F_{\rm min}$ -- the minimum flux of the hot component;  \\
$\Delta F = F_{\rm max} - F_{\rm min} $ -- peak-to-peak amplitude of the flickering; \\
$F_{\rm av}$ -- the average flux of the hot component:    
\begin{equation}
  F_{\rm av} = \frac{1}{N}  \sum_{i=1}^{N} F_i; 
\end{equation}
$F$$_{\rm fl}$ -- the average flux of the flickering, F$_{\rm fl}=$F$_{\rm av}
-$F$_{\rm min}$; \\ 
and the absolute rms amplitude of variability  (the square-root of the light-curve variance):
\begin{equation}
  \sigma_{\rm rms_0} =  \sqrt{ \frac{1}{N-1}   \sum_{i=1}^{N} (F_i - F_{\rm av} )^2  },
\end{equation}
where $N$ is the number of the data points in the run (as  used in
Uttley, McHardy \& Vaughan, 2005).  
Subsequently,
we subtract the contribution expected from measurement errors
\begin{equation}
  \sigma_{\rm rms} =  \sqrt{ {\sigma_{\rm rms_0}}^2 - {\sigma_{\rm err}}^2  } ,
   \label{eq.err}
\end{equation}
where  $\sigma$$_{\rm err}$ is the mean observational error.

Following King et al. (2004), we calculate rms flux:
\begin{equation}
  \sigma  = \frac{1}{N} \sqrt{  \sum_{i=1}^{N} (F_i - F_{\rm av} )^2  }.
\end{equation}
We correct $\sigma$ for the observational errors in a similar way as equation~\ref{eq.err}.
The corrections of $\sigma$ and  $\sigma$$_{\rm rms}$ for the measurement errors are small, in the range 1--4\%.  

In Fig.~\ref{fig.quant}  are plotted the flickering quantities $F_{\rm max}$, $F_{\rm min}$ and $F_{\rm fl}$ versus  the average flux of the  
hot component. The individual errors are also indicated (in most cases
they are less than or equal to the size of the symbols).
Although we expect them to be connected,   
it is not clear a priori how  the different quantities 
depend on each other.   
Least-squares fit to data (taking into account the errors of the
individual points) 
in Fig.~\ref{fig.quant} give:  
\begin{eqnarray}  
F_{\rm max}= 1.31\,(\pm 1.64) + 1.161\,(\pm 0.013)\,\,F_{\rm av},   
\label{eq.1ot3}\\ 
F_{\rm min}= -0.90\,(\pm 1.17) + 0.857\,(\pm 0.009)\,\,F_{\rm av},\;\;{\rm and}  
\label{eq.2ot3}\\    
F_{\rm fl} = -0.33\,(\pm 0.19) + 0.135\,(\pm 0.002)\,\,F_{\rm av},  
\label{eq.3ot3}
\end{eqnarray}  
where the units for all quantities are $10^{-14}\,{\rm erg}\,{\rm
  cm}^{-2}\,{\rm s}^{-1}\,\AA$.
The relations (equations~\ref{eq.1ot3}, \ref{eq.2ot3}, and \ref{eq.3ot3}) are
very similar to those calculated for  
the recurrent nova T~CrB  (Zamanov et al. 2004).

\section{Results}

\subsection{Relationship between $\Delta F$  and $F_{\rm av}$}
\label{FliAmp}

\begin{figure*}    
   \vspace{11.5cm}   
   \includegraphics{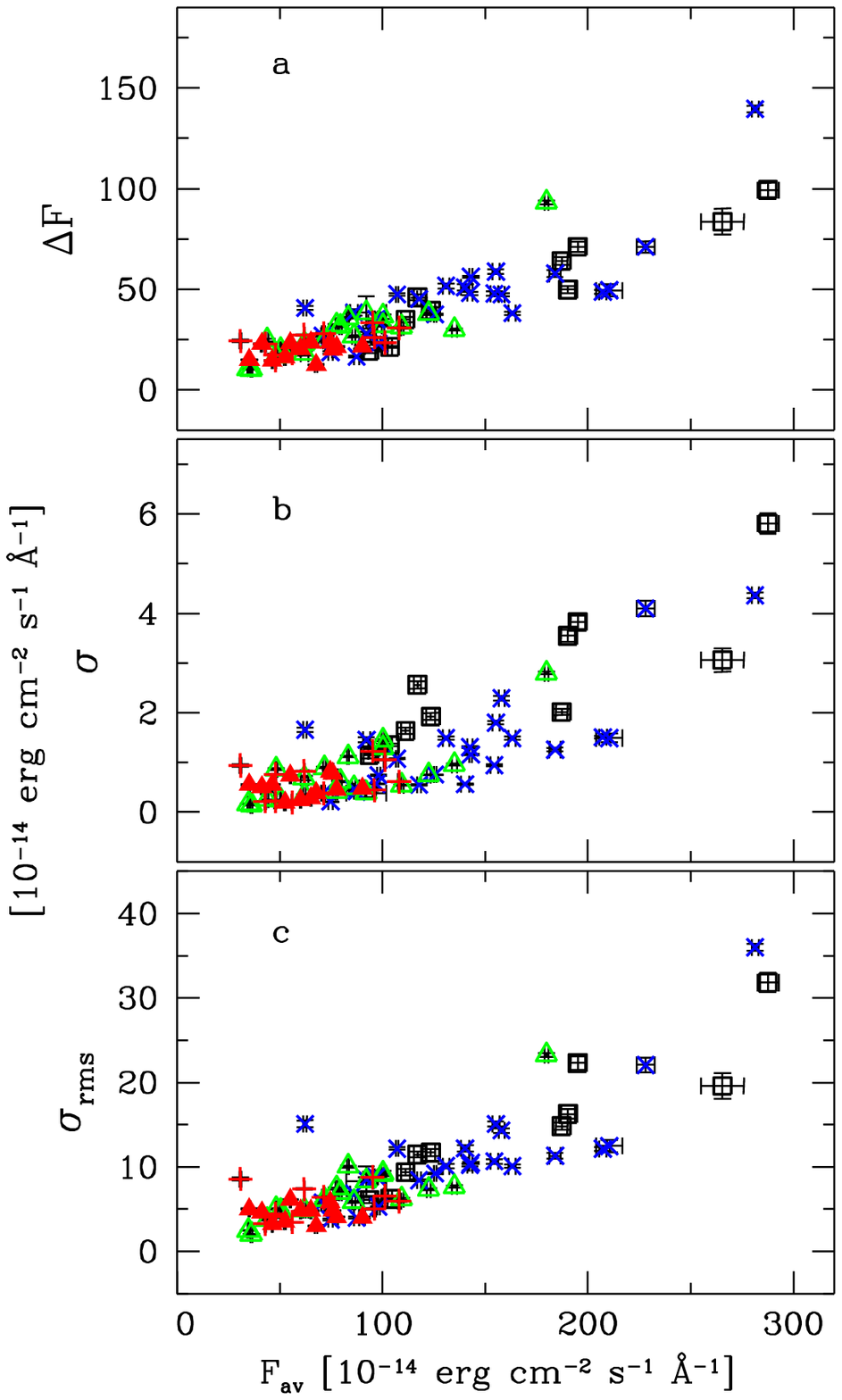}      
   \includegraphics{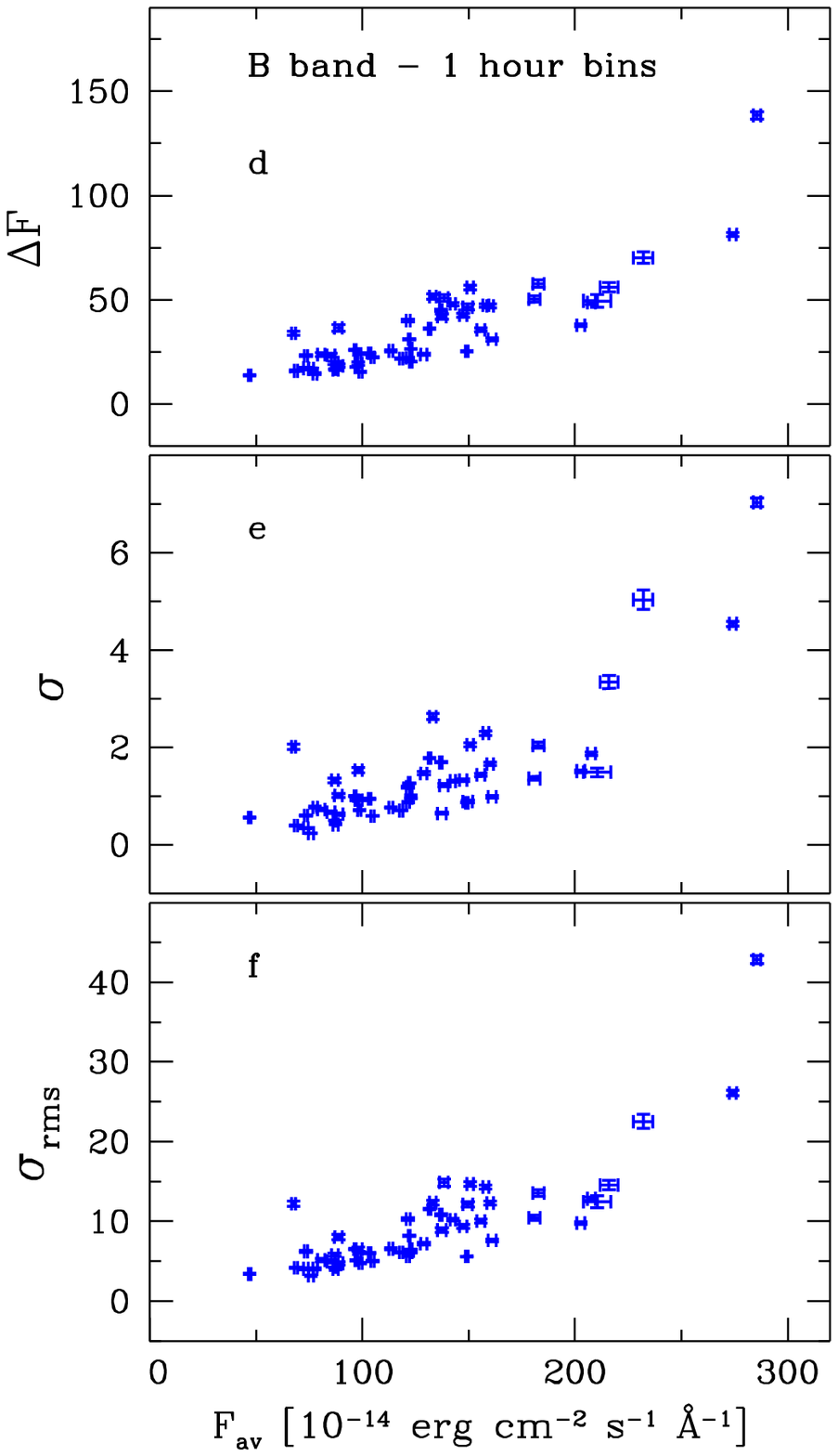}      
   \includegraphics{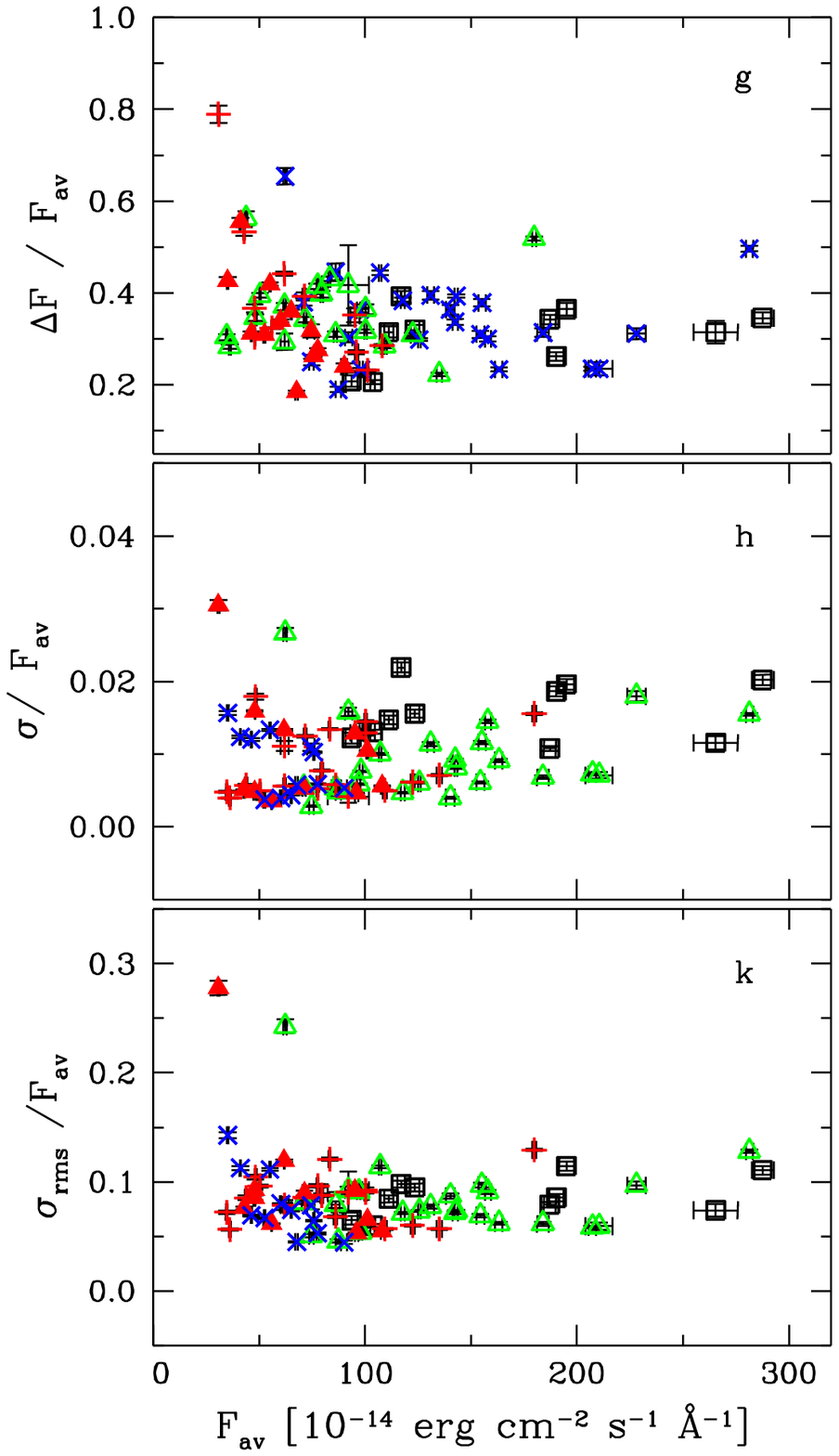}       
   \caption[]{RS~Oph: amplitude of the flickering versus the dereddened average flux of the hot component 
    in $UBVRI$ band.   
    The symbols are the same as in Fig.\ref{fig.quant}. 
    In (a), (b), and (c) panels are  plotted amplitude $\Delta F$, $\sigma$, and $\sigma_{\rm rms}$ respectively. 
    In (d), (e), and (f) panels are plotted $\Delta F$, $\sigma$, and $\sigma_{\rm rms}$ for 1 h bins.
    In (g), (h), and (k)  panels is the normalized  variability.
   \label{fig.ampl} }         
 \end{figure*}

From the amplitude of the flickering versus the average flux plotted
in Fig.~\ref{fig.ampl}(a), 
it is clear that $\Delta F$ increases with $F_{\rm av}$. We performed Pearson's  correlation test  and Spearman's (rho) rank
correlation test.  The results of these tests (correlation coefficient
and $p$-value) are summarized in 
Table~\ref{Tab1}, where the first column lists the band(s) used, the
second column the number of observations, the third 
the power-law index and its error, the fourth Pearson's
correlation coefficient, and 
the fifth and sixth Spearman's correlation coefficient and its 
significance ($p$-value).  The correlation between  $\Delta F$ and $F_{\rm av}$ is highly significant 
($p << 0.001$) even for a single band. Because the significance depends on the number of data points, we
obtain higher significance when we use more data, achieving $p \sim 10^{-20}$, when we use all 73 light curves in $UBVRI$ bands.

Our runs had durations from 40 min to 3 h, and the differing durations could in principle affect 
the behaviour of the flickering amplitude (because for red noise, rms amplitude depends on the time interval
over which it is measured). To check that the differing light-curve durations were not influencing our results, 
we divided our $B$-band data into 1 h segments. We re-calculated sigma and amplitude.  
The result was similar to that obtained with the complete light curves (see Fig.~\ref{fig.ampl} d, e, f, and Table~\ref{Tab1}). 

\begin{figure}    
   \vspace{6.0cm}   
   \includegraphics{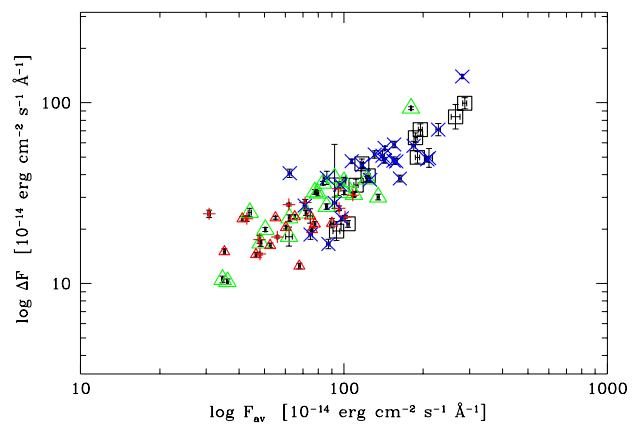}      
   \caption[]{RS~Oph: amplitude of the flickering versus the average flux in 
    $UBVRI$ bands, on a logarithmic scale.  For the majority of points, the
    individual errors are less than or equal to the size of the symbols.
    The symbols are the same as in Fig.\ref{fig.quant}. 
    }
   \label{fig.5}    
 \end{figure}

Looking for a dependence of the type $\Delta F \propto F^k_{\rm av}$, we fit the data to a straight line in log--log space, 
taking errors into account (Fig.~\ref{fig.5}).  The results for the power-law index $k$ are
summarized in the third column of Table~\ref{Tab1}.  For the power-law
index $k$, the calculated values are in the range $1.02 \le k
\le1.32$. In general, $k = 1.02 \pm 0.04$, which is derived from all
of the observations, 
should provide a better measure of power-law index.

\subsection{Relationship between rms variability and average flux}
\label{Flisig}

The rms variability ($\sigma$,
$\sigma$$_{\rm rms}$) 
and $F_{\rm av}$  
are also strongly correlated. 
Using all five bands for the correlation between  $\sigma$ and
$F_{\rm av}$ (Fig.~\ref{fig.ampl}b),   
we calculate a correlation coefficient of 0.80 and a significance of
$p = 3 \times 10^{-14}$.
For $\sigma$$_{\rm rms}$ and $F_{\rm av}$ (Fig.~\ref{fig.ampl}c), we calculate a
correlation coefficient of 0.84 and a significance of
$p = 6 \times 10^{-17}$. 
The relationship between the rms variability and the mean flux
is consistent with a linear dependance.  

In Figs.\ref{fig.ampl}(g), (h), and (k)
(right-hand panels), we plot the fractional
variability, i.e., the variability normalized 
by $F_{\rm av}$. There are a few deviating points (the deviation is 
most clearly visible in Fig.3k), which  
are due to 20150815 run (see Sect.~\ref{2012...08...15} for details).
We estimate mean values  $< \sigma_{\rm rms}  / F_{\rm av} > = 0.08 \pm 0.02 $ 
and   $<\sigma / F_{\rm av}>  = 0.010 \pm 0.004$. Our data show that the fractional rms variability 
remains approximately constant, despite significant flux changes ($\sim$factor of 4.4 in B band).
Following King et al. (2004), 
$\sigma / F_{\rm av}$
depends on 
the Shakura \& Sunyaev (1973) $\alpha-$viscosity parameter in the
accretion disc (see section~3.2.2  of King et al. 2004).
For RS~Oph, we calculate 
$\sigma / F_{\rm av} \sim 0.01$, which corresponds to $\alpha \le 0.006$.

\subsection{Flux distribution}
\label{lognor}
A characteristic of the X-ray variability in X-ray binaries and active galaxies
is the log-normal flux distribution (Uttley et al. 2005). This
distribution is also found
in the $Kepler$ observations of V1504~Cyg and KIC~8751494 (Van de Sande, Scaringi \& Knigge 2015).

\begin{figure}    
   \vspace{14.0cm}   
   \includegraphics{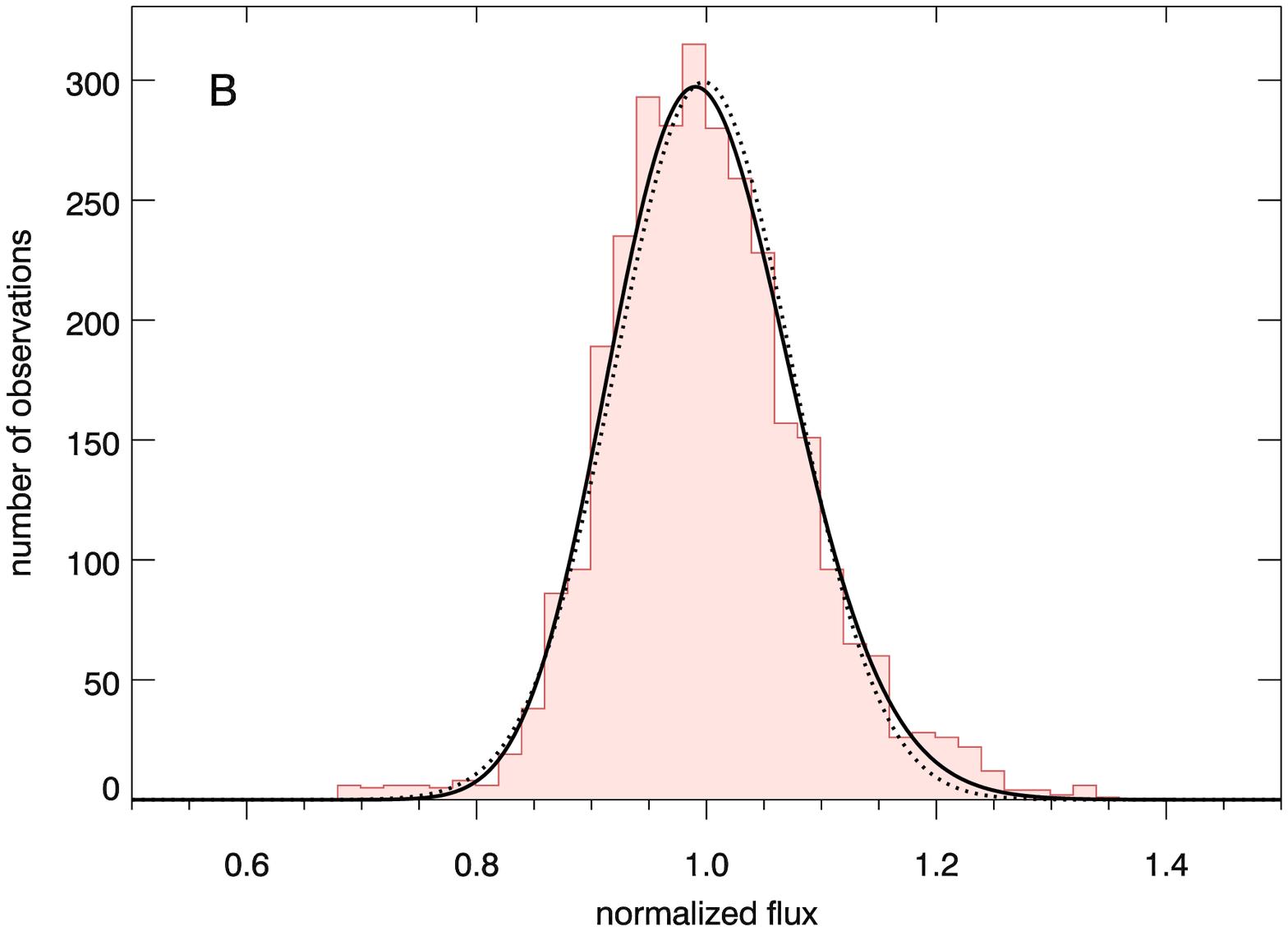}      
   \includegraphics{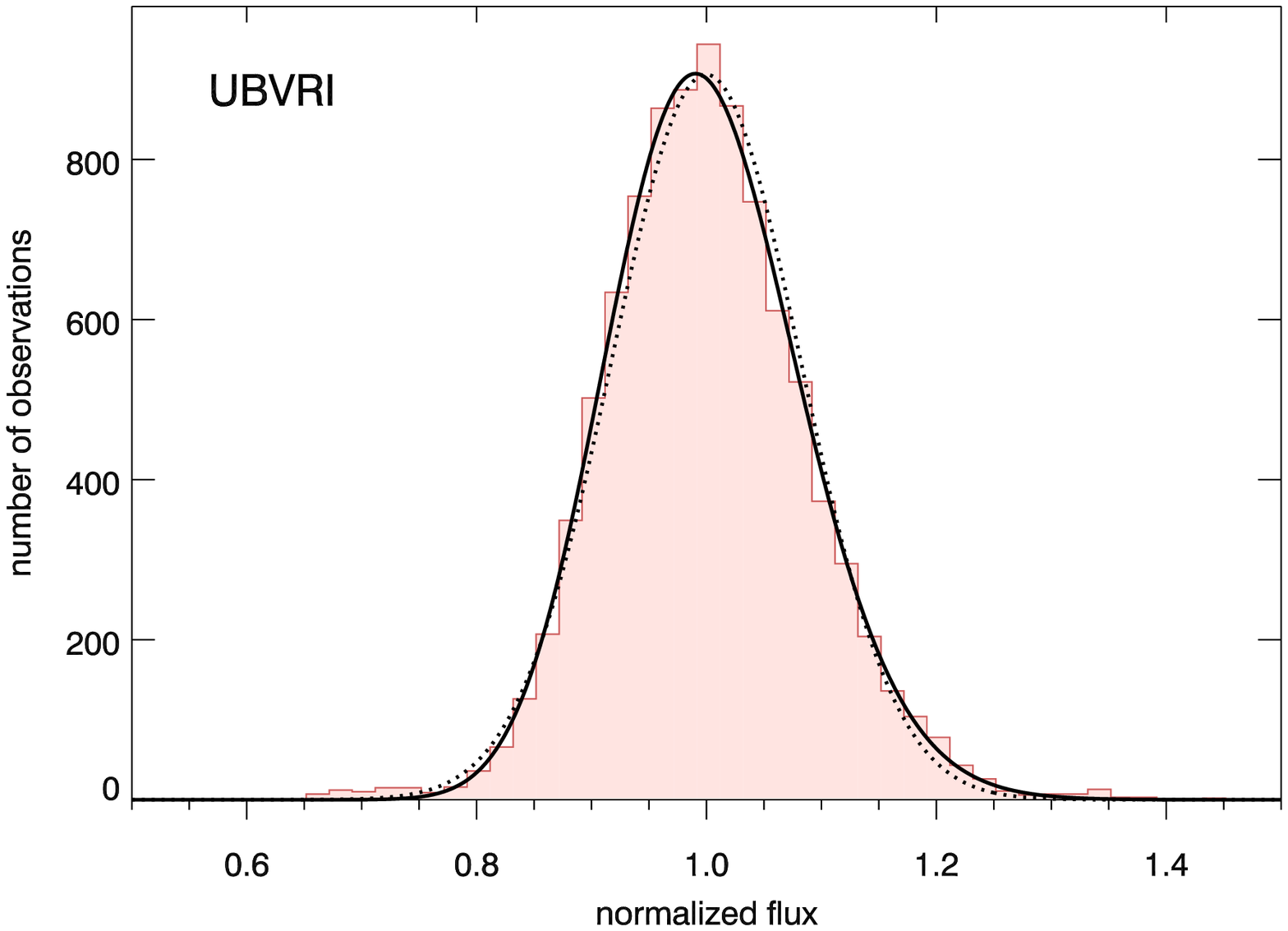}      
   \caption[]{Flux distribution of RS~Oph, $B$-band data (upper panel), and all
    $UBVRI$ data (lower panel). Scott's normal reference rule was used to determine the binsize of 0.02, because it
    is optimal for random samples of normally distributed data. 
    Log-normal distribution fits are drawn with solid lines, the dotted lines represent
    Gaussian distribution fits.
    }
   \label{fig.lognor}      
 \end{figure}

To check whether RS~Oph exhibits a similar behaviour, we performed
both Gaussian and log-normal fits to the flux distribution of the
$B$-band data and to all of the 
$UBVRI$ data (Fig. \ref{fig.lognor}).
For the $B$-band data the fit to the normal (Gaussian) distribution gives  reduced
$\chi2 = 4.53$ (centre of the distribution 0.9977, standard deviation  0.0768, degrees of freedom 31).
The log-normal distribution fit with lognormal centre of 0.9997 and
lognormal standard deviation of 0.0792 gives reduced $\chi2 = 3.06$  (degrees of freedom 31).
For $UBVRI$ sample the normal distribution fit (centre 0.9997, standard deviation 0.0822) gives 
reduced $\chi2 = 4.82$ (degrees of freedom 37), while  the lognormal distribution fit 
gives reduced $\chi2 = 2.51$ (lognormal centre of 1.0008 and
lognormal standard deviation of 0.0835, degrees of freedom 37).


The results show that the log-normal distribution provides a better  fit to the flux distribution  
of RS~Oph in the optical bands because it better accounts for the skew. 

\section{Discussion}

The flickering of RS~Oph disappeared after the 2006 outburst 
(Zamanov et al. 2006), indicating that the accretion disc was destroyed
by the blast wave from the nova.  
Photometric data of Worters et al. (2007) showed evidence of the resumption of
optical flickering, indicating reestablishment of accretion by day 241
of the outburst.  
Most of the present data were obtained after the reappearance of the
flickering.  
That the data obtained  before and after the outburst (1997--1998) 
and (2009--2013) 
exhibit the same behaviour 
confirms
that the behavior  
we see persists over long time periods (more than a decade). 

\subsection{Light curve from 2012 August 15}
\label{2012...08...15}
Kundra, Hric, \& G{\'a}lis (2010) performed  wavelet analysis of the
flickering of RS~Oph and 
found two different sources of flickering, the first with
amplitude 0.1 mag and frequencies of 
60--100 cycles d$^{-1}$,  and the second with amplitude 0.6 mag and
frequencies of less than 
50 cycles d$^{-1}$. These frequencies are also visible in most of our
observations  
(Fig.~\ref{fig.example}, left-hand and right-hand panels). 
However, they are not visible in our  20120815 run (see Fig.\ref{fig.example}, mid panel).
The light curve of RS~Oph obtained on 2012 August 15 shows smooth 
variations. It resembles the $B$-band light curve of CH~Cyg obtained on 1997 June 9 (see fig. 1 of Sokoloski \& Kenyon 2003b), during which time Sokoloski \& Kenyon (2003a) 
suggested that the inner disc was disrupted due to the launch of a jet.
The fractional rms variability during this run 
was $\sigma$$_{\rm rms}$ $/ F_{\rm av} = 0.23$. 
This value deviates considerably from 
that of other runs (see Fig.~\ref{fig.ampl}f), 
indicating that at this moment the accretion disc was in an unusual state.

\subsection{Amplitude of flickering}

We detect (Sect.\ref{FliAmp}) correlation between $\Delta F$ and
$F_{\rm av}$. Such a correlation has 
already been detected in few other binaries. Analysing the $U$-band
flickering of  
the symbiotic star CH~Cyg in 1974--1989, Mikolajewski et al. (1990)
found that the amplitude of the flickering in $U$-band is a power-law
function of the flux,  
$\Delta F  \sim F^k$, with $1.40 < k < 1.45$.
For the recurrent nova  T~CrB,  Zamanov et al. (2004) estimated $k=1.03--1.09$,
and for the nova-like CV star KR Aur, Boeva et al. (2007) gave $k=0.70-0.75$.

{\footnotesize
\begin{table} 
\caption{Correlation analysis for the relationship between amplitude of flickering and the average flux of the hot  component. 
 In the table are given the band(s) used, $N_{\rm obs}$ (the number of light curves), 
 power-law index $k$, Pearson and Spearman rank correlation coefficients, and their significance.}
\begin{center}
\begin{tabular}{crccclcllrrr}
\hline

Bands     &$N_{\rm obs}$&        $ k $     &  $r_{\rm P}$  &       $r_{\rm S}$        &      $p$--value       &   \\
\\
 $UBVRI$     &  73    & $1.02 \pm 0.04$ &   0.88  &        0.86       & $2 \times 10^{-23}$ &   \\
 $B    $     &  23    & $1.27 \pm 0.12$ &   0.83  &        0.80       & $5 \times 10^{-6}$  &   \\  
$B$ 1 h bins  &  52    & $1.32 \pm 0.08$ &   0.84  &        0.83       & $3 \times 10^{-14}$ &   \\
 $V    $     &  19    & $1.11 \pm 0.08$ &   0.88  &        0.80       & $2 \times 10^{-5}$  &   \\  
 $UB   $     &  33    & $1.31 \pm 0.10$ &   0.87  &        0.87       & $6 \times 10^{-11}$ &   \\  
 $BV   $     &  42    & $1.05 \pm 0.05$ &   0.89  &        0.87       & $5 \times 10^{-14}$ &   \\   
 $UBV  $     &  52    & $1.06 \pm 0.05$ &   0.89  &        0.88       & $6 \times 10^{-18}$ &   \\

\hline
\end{tabular}
\end{center}
\label{Tab1}
\end{table}
}

Our results for RS~Oph point to a value  of  $k = 1.02 \pm 0.04$,  not surprisingly similar to the 
value derived in T~CrB. There are many similarities between these two
'sister' systems, e.g.,  
they both: (1) are recurrent novae; 
(2) harbour very massive WDs; and
(3) accrete at similar rates -- RS~Oph:  $2 \times 10^{-8}$ M$_\odot$~yr$^{-1}$  (Nelson et al. 2011), 
T~CrB: $2.5 \times 10^{-8}$ M$_\odot$~yr$^{-1}$  (Selvelli \& Gilmozzi 1999). 
  
An increase in brightness (and $F_{\rm av}$) is usually due to an increase
in the mass accretion rate. 
If the flickering is coming from the boundary layer between 
the accretion disc and the WD (innermost part of the
accretion disc), we expect the correlations 
between $F_{\rm av}$ and other quantities ($F_{\rm min}$, $F_{\rm max}$, $F_{\rm fl}$, $\Delta F$, 
$\sigma$$_{\rm rms}$) 
to be connected with the response (changes in the structure and/or 
the size) of the boundary layer to the changes in $\dot M_{\rm acc}$.  
If the flickering is coming from a hotspot (outer part of the
accretion disc), we expect the correlations 
to be associated with the size (mass) of the accreting blobs.

Dobrotka et al. (2010) analysed V-band photometry of the aperiodic
variability in T CrB. By applying an prescription for
angular momentum transport in the accretion disc, 
they developed a method to simulate the statistical distribution
of flare durations under the assumption that the aperiodic variability
is produced by 
turbulent elements in the disc. Furthermore, the simulated light curves (Dobrotka et al. 2015) 
exhibit the typical linear rms--flux relation and log-normal distribution.

In CVs, many statistical properties of the
flickering are explained with the fluctuating accretion disc model in
which variations in the mass transfer rate through the disc are
modulated on the local viscous time-scale and propagate towards the
central compact object (Scaringi 2014).
Alternatively, 
Yonehara, Mineshige \& Welsh (1997) proposed a model in which light
fluctuations are produced by occasional  
flare-like events and subsequent avalanche flow in the accretion disc
atmospheres.  
Flares are assumed to be ignited when the mass 
density exceeds a critical density. 
In this model, the correlations  
could be connected with the size of the element, where the  density exceeds a critical density.

\subsection{Analogy with  accreting black holes}

One feature of the broad-band  X-ray variability of accreting black holes
is the so-called rms-flux relation, which is a linear relationship
between the absolute 
rms amplitude of variability and the flux, such that sources become 
more variable as they get brighter. 
Uttley \& McHardy (2001) found this  relation  in  Cyg~X-1 (black hole mass 14.8 M$_\odot$; Orosz et al.~2011) 
and in the accreting millisecond pulsar SAX J1808.4-3658 (neutron star mass $\approx$ 2 M$_\odot$; Wang et al.~2013).
Heil \& Vaughan (2010) reported the detection of this relation 
in the ultraluminous X-ray source  NGC 5408 X-1 (black hole mass 100--1000 M$_\odot$).


A similar linear relationship between the rms
variability amplitude and the mean flux 
was discovered in the CVs MV Lyrae
(Scaringi et al. 2012), V1504~Cyg, and KIC8751494 (Van de Sande et al. 2015).  
In MV~Lyr, the WD mass is 
$\approx$ 0.72 M$_\odot$, and  the mass transfer rate
ranges from $3\times 10^{-13}$ $M_\odot$~yr$^{-1}$ (Hoard et al. 2004)
to $8.5 \times 10^{-10}$ $M_\odot$~yr$^{-1}$ (Echevarr{\'{\i}}a  1994) and
$2 \times 10^{-9}$ $M_\odot$~yr$^{-1}$ (Godon \& Sion 2011).
The observations reported here indicate that a similar relationship
exists in the case of RS~Oph,  
although in RS~Oph the WD mass is close to the Chandrasekar
limit (Brandi et al. 2009) and 
the  mass accretion rate is a few orders of magnitude higher 
$2\times10^{-7} - 2 \times 10^{-8}$ $M_\odot$~yr$^{-1}$  (Osborne et
al. 2011; Nelson et al. 2011). 

The rms--flux relation remains an enigmatic observational feature of
accreting compact objects, but it clearly 
contains information about the dynamics of the infalling material. 
The similarities between the behaviour of the optical flickering
amplitude in WD accretors  
and the X-ray variability of accreting black holes indicate
that similar processes may produce the short-term variability 
in the accretion flows around WDs, stellar mass black holes,
and super--massive black holes.  
The deviating points (like 20120815) could help us to better
understand the physical processes producing the short-term
variability.  


\section{Conclusions}

We present observations of the flickering variability of the recurrent nova RS~Oph at quiescence 
in the optical $UBVRI$ bands. We find a  highly significant correlation
between the flickering amplitude and  
the average flux of the hot component. We estimated the relation between
the average flux of the hot component and various flickering quantities ($F_{\rm min}$, $F_{\rm max}$, $F_{\rm fl}$, $\Delta F$, 
$\sigma$$_{\rm rms}$).
The amplitude--flux $(\Delta F$ versus $F_{\rm av})$ and rms--flux $(\sigma$ versus $F_{\rm av})$ as well as the other 
relations contain information about the infalling material 
in the accretion disc and should be useful to test the theoretical models of flickering.

\section*{Acknowledgements}

This work is partially  supported by the OP 'HRD', ESF and Bulgarian  Ministry of Education and Science
(BG051PO001-3.3.06-0047). JLS acknowledges support from NSF grant AST-1217778. 




 															     
\bsp

\label{lastpage}

\end{document}